\documentclass[preprint,aps,prd,showpacs]{revtex4}
\usepackage{psfig}
\begin{document}
\renewcommand{\thesection}{\arabic{section}}
\renewcommand{\thesubsection}{\arabic{subsection}}
\title{On the Low Surface Magnetic Field Structure of Quark Stars}
\author{Nandini Nag, Sutapa Ghosh, Roni Saha and Somenath Chakrabarty$^\dagger$}
\affiliation{
Department of Physics, Visva-Bharati, Santiniketan 731 235\\ West Bengal, India,
$^\ddagger$E-mail:somenath.chakrabarty@visva-bharati.ac.in}
\begin{abstract}
Following some of the recent articles on hole super-conductivity and related 
phenomena by Hirsch \cite{H1,H2,H3}, a simple model is proposed to explain the 
observed low surface magnetic field of the expected quark stars. It is argued 
that the diamagnetic moments of the electrons circulating in the electro-sphere
induce a magnetic field, which forces the existing quark star magnetic flux 
density to become dilute. We have also analysed the instability of 
normal-superconducting interface due to excess accumulation of magnetic flux 
lines, assuming an extremely slow growth of superconducting phase through a 
first order bubble nucleation type transition. 
\end{abstract}
\maketitle
\section{Introduction}
More than two decades ago, Witten had conjectured in an outstanding article 
\cite{EW} (see also \cite{ARB}) that a flavor symmetric quark matter at zero 
temperature and at zero pressure would be energetically the most stable 
configuration. This exotic phase is known as the strange quark matter (SQM). 
Since then a large number of articles have been published on the
theoretical studies of the
properties of bulk SQM as well as droplets (known as strange-lets) of SQM 
\cite{SQ1,SQ2,SQ3}. In those investigations various kinds of confinement models
were used. It has further been shown that SQM phase can exist in compact 
stellar objects, known as strange quark stars or strange stars \cite{SS1,SS2}. 
The finite size strange-lets are expected to be formed both in ultra-relativistic 
heavy ion collisions and also during QCD phase transition in the early universe 
\cite{RC,EU}. The strange-lets expected to have formed in the early universe, 
are also called the strange nuggets, believed to be the relics from primordial 
quark-hadron phase transition. These relics are also one of the viable candidates for 
baryonic dark matter. Inside the strange stars, three light flavors: up ($u$), 
down ($d$), strange ($s$), along with small quantity of electrons are in 
$\beta$-equilibrium. The presence of electrons make the system electrically 
charge neutral. So far the gross properties of such compact objects are 
concerned, it is almost impossible to distinguish strange stars from the 
conventional neutron stars. Except, it is believed that these compact objects, 
which are expected to be strange stars, are fast rotating (milli second or 
sub-milli second pulsars) and the surface magnetic field is quite low (
$\leq 10^8$G) \cite{OSS}. It is further assumed that strange stars are very 
old and also extremely cold objects, formed by the accretion of matter from 
the companion binary counter part. However, for a strange star, the 
compositional structure in the microscopic scale is extremely complex in 
nature. Since the density at the core region is very high ($3-4$ times normal 
nuclear density), the corresponding quark matter is almost flavour symmetric. 
As one goes radially outward, the density of matter decreases and since 
s-quarks are much more heavier than $u$ and $d$-quarks (current masses for $u$ 
and $d$-quarks are $5-10$MeV, whereas, for $s$-quark, it is $\approx 150$MeV), 
in this region, it is therefore energetically not favorable to produce enough 
$s$-quarks through weak processes. The abundance of $s$-quark is therefore 
steadily decreases as one moves towards the surface region. The charge 
neutrality of quark matter near the outer core or the crustal region is 
therefore mainly maintained by electrons instead of strange quarks. The density
of electron is negligibly small at the core and is maximum at the crustal 
region. Of course it is not very high ($\leq 10^{-4}$ times normal nuclear density). 
These electrons are bound to the positively charged quark matter by 
electromagnetic force (of which the Coulomb force is the
non-relativistic scenario) and are allowed to move freely across the
strong Coulomb field. Because of this strong electromagnetic force, they can 
not move far away from the quark matter surface. The electron gas at the 
surface of strange stars, form a very thin layer of width a few thousand Fermi,
known as electro-sphere. Therefore, the gross compositional structure of a 
strange star is very simple: a positively charged quark matter region covered
by a thin electro-sphere. Since the strange quark matter is energetically more
favorable, in the case of strange stars, it is unlikely to have a crust of
dense iron like matter, which is expected to be present in neutron
stars. Further, depending on the internal temperature and density of 
strange star, the quark matter may be in normal phase or in super-conducting 
state or in color flavor locked phase (CFL) \cite{RG}. The CFL phase at the core
region is expected if the density is extremely high. Since the kinetic energies
of the electrons are much larger than the corresponding super-conducting gap 
energy, they never show super-conductivity, either inside the star or at the 
electro-sphere. Which actually means that the critical temperature for
superconducting transition of electron is much less than the
corresponding critical temperature for quark matter.

In this article we shall propose a mechanism by which very low surface
magnetic field of strange stars can be obtained. We have assumed a
type-I superconducting transition and consider two
possible flux expulsion processes. We shall show that extremely slow expulsion
mechanism leads to an instability at the normal-superconductor interface
of quark matter. In this article we have also analysed the interface instability. It is
further noticed that by a slow expulsion process, at several points on
the interface, since the strength of magnetic field becomes $> B_c$, the
critical strength to destroy type-I superconductivity of quark matter, the growth of
superconducting quark matter phase will stop abruptly. On the other
hand, if it is an extremely fast process, we ultimately get
a stable quark star of very low surface magnetic field. 

The paper is organized in the following manner: In the next section we
shall compare a strange star with a giant atom. In this section we shall
discuss qualitatively the physical processes which may take place at the 
surface of strange star and compare them with a giant ultra-heavy atom. In 
section 3. we shall make a comparative study of superconducting quark
matter in strange star and the theory of hole superconductivity. The
comparison is again purely qualitative. The mathematical formalism will be
given in section 6. We shall discuss the mechanism
of slow flux expulsion and give a mathematical formalism in sections 5
and 6 respectively. In the last section we have given the conclusion of
the present work.

\section{A Giant Atom Model}
In this section we shall try to explain the possible low magnetic
surface structure of quark stars. In this model calculation, we assume a
prompt phase transition to superconducting phase and the strange star is treated 
as an equivalent giant atom with positively charged quark matter (along with 
some admixture of electrons) as the nucleus and the electrons in the 
electro-sphere are treated as orbital electrons. Unlike normal atoms, here size of the nucleus is $\gg$ than the volume occupied by the electrons in the 
electro-sphere. Now it is very easy to show from the $\beta$-equilibrium and 
charge neutrality conditions that the net positive charge content of the quark 
matter nucleus is $\gg 172$, the critical value of $Z$, the atomic number of a 
typical super-heavy Dirac-atom. At the quark matter surface, the vacuum will therefore 
become unstable, and spontaneous $e^+e^-$-pairs will be created if there are 
unoccupied energy shell \cite{D1} (see also \cite{JM}).

Now for a many body fermion system, the microscopic theory of superconductivity
suggests that if the interaction favors the formation of pairs at low 
temperature, the system may undergo a phase transition to a super-fluid state. 
In astrophysics, this is expected to occur in the dense neutron matter of 
cold neutron stars. Whereas the small percentage of protons ($\sim 4\%$) inside 
neutron stars undergo a transition to type-II superconducting phase 
\cite{SF1,SF2}. In the case of a many body system of fermions 
the well known BCS theory is generally used to study the super-conducting 
properties due to fermion pairing. One fermion of momentum $\vec p$ and 
spin $\vec s$ combines with another one of momentum $-\vec p$ and spin 
$-\vec s$ and form a Cooper pair. In the case of type-I electronic 
super-conductors the coupling is mediated by the electron-phonon interaction. 
In the case of quark matter, however, the basic quark-quark interaction at 
large distance favors the formation of Cooper pairs. In the case of
quark matter, since the force is mediated via gluons, it gives
rise to what is known as color superconductivity. For a highly degenerate 
fermion system, which is true for strange star matter, the pairing takes place 
near the Fermi surface. The other important condition that must be satisfied 
for the formation of Cooper pairs is that the temperature ($T$) of the system 
should be much less than the super-conducting energy gap ($\Delta$). In the 
case of strange stars, since the Fermi levels are not identical for $u$, $d$ 
and $s$-quarks, only same type of quarks can form Cooper pairs at the
Fermi surface and give rise 
to color super-conductivity \cite{BL}. It is interesting to note that the 
electrons, whose density is extremely low compared to quark matter, may also 
be treated as highly degenerate relativistic plasma, but are unlikely to form 
Cooper pairs.

In the next section we have assumed a type-I super-conducting phase of 
quark matter within the strange stars and develop a mechanism by 
which the expelled magnetic field is reduced at the electro-sphere. 
Since we are not investigating any of the properties of super-conducting quark 
matter, rather, we are interested to study some of the important features of 
normal electron gas layer at the surface region / inside the
electro-sphere, which are essential to have low surface magnetic field,
then instead of standard relativistic version of 
BCS theory \cite{BL}, here, we have followed the interesting idea of 
"theory of hole super-conductivity", proposed by Hirsch in a series of articles
\cite{H1,H2,H3,H4,H5}. In the case of strange stars the charge separation or 
the charge asymmetry takes place by the combined effect of charge
neutrality and $\beta$-equilibrium conditions; completely different physical
mechanism. Further, the properties of both positively charged quark matter
inside the star and the electrons in the electro-sphere are just opposite to
the predicted nature of hole states and the electrons for the hole 
super-conductor. In fact, we have noticed that the theory of hole 
super-conductivity is not applicable for super-conducting quark matter of 
strange stars. The reason is the strong interaction which bind the
quarks within the strange stars. 

\section{The Hole Superconductivity and Compact Strange Stars}
To develop a mechanism of strange star magnetic field suppression at the
electro-sphere, we consider the charge asymmetric strange star as an
equivalent hole super-conductor. We have noticed, that there are a lot of 
similarities and also dissimilarities between the laboratory samples and the 
strange stars within the theory of hole super-conductivity. In the case of 
laboratory superconducting samples, according to the theory proposed by Hirsch,
the inner part is positively charged holes in normal phase. The electrons at 
the surface region are super-conducting. The surface 
layer is bound to the inner region by strong electromagnetic force. In the case
of strange stars, the inner region is positively charged quark matter. But 
unlike the laboratory superconductors, it is in super-conducting phase. 
Whereas, the outer layer, the electro-sphere, which is a degenerate electron 
gas, is in normal phase. The force between these two regions is again 
electromagnetic in nature. Going a step further, we assume following the
model by Hirsch, that the electrons in 
the electro-sphere gyrate about the magnetic lines of force, expelled from the 
super-conducting interior of the strange star. The origin of such orbital 
motion of these normal electrons in electro-sphere about the magnetic flux lines is 
the well known Lorentz force. This orbital motion of electrons will generate a 
magnetic field at the surface of the strange star. According to Lenz's law, 
the induced field must be in the direction opposite to that of existing one. 
At the surface or in the electro-sphere, therefore, the magnetic field from the tiny magnets 
produced by the gyration of electrons about the existing lines of forces reduce
the strange star magnetic field by diamagnetic effect. Therefore, unlike the 
hole super-conductor, in the case of strange stars, the magnetic field 
expelled from the super-conducting quark matter are suppressed by the electron 
gas, which is in normal phase. By the repulsive diamagnetic action, there will be an effective dilution of magnetic flux lines 
inside the electro-sphere. The dilution of magnetic flux will naturally 
increase the size (width) of the electro-sphere. Analogous to the phenomenon 
of frozen-in magnetic field, here, the "{\it{frozen-in electron gas}}" will be 
pulled away by the flux lines. The increase in size of the electro-sphere will 
depend both on the repulsive magnetic force and the attractive coulomb force 
and in equilibrium configuration, these two will balance each other. The width 
of the electro-sphere of a typical quark star of surface magnetic field $\leq
10^8$G can be a few tens of km, which is at least an order of
magnitude less than the radius of the light cylinder, $\sim 100$km, for a
milli second pulsar. Now this increase in size of the electro-sphere will 
reduce the electron density within the system. The relation between the induced
field and the existing field is given by eqn.(39). Further, because of the
motion of electrons around the quark matter nucleus, the actual trajectory of 
an electron in the electro-sphere is more or less like a closed helical spring 
produced by the motion of circular orbit along the lines of force. Since the 
magnetic field strengths at the poles are very high, the electrons in the 
helical trajectory will be reflected back from these region. The two ends will 
therefore behave like magnetic mirrors. It is also obvious from eqn.(39)
that the radius of gyration is larger at the equatorial region compared
to the polar values. Since the diamagnetism of the electrons
inside the electro-sphere is caused by Lenz's law, the sign of 
$\vec L.\vec p/\vert \vec L. \vec p\vert$, which may be called as the effective 
"orbital-helicity" will change sign after each reflection by the
magnetic mirrors at the poles. Here $\vec p$ and $\vec L$ are
respectively the linear momentum and the angular momentum of the
electrons. Further, when a balance between the existing 
field and the induced field will be established, the electrons within each 
half of the electro-sphere will under go steady helical motion.

\section{Slow Expulsion of Magnetic Flux Lines}
In this section we shall give a qualitative picture of very slow magnetic flux
expulsion from the super conducting region. It has been shown in
ref.\cite{BL} in a relativistic version of BCS theory, that if a normal quark 
matter system undergoes a superconducting phase transition, the newly produced
quark matter phase will be a type-I superconductor. They have also shown in 
that the critical magnetic field to destroy such pairing is  $\sim  10^{16}$G
for $n\sim 3-4n_0$, with $n_0=0.17$fm$^{-3}$, the normal nuclear
density. This magnetic field strength is indeed much larger than the typical
pulsar magnetic field. The corresponding critical temperature  is
$\sim 10^9-10^{10}$K, which is again high enough for strange stars,
which are expected to be extremely cold objects. In this section, instead of 
investigating the superconducting properties of quark matter inside strange 
stars, we shall give a possible mechanism of flux expulsion by Meissner
effect, assuming that the growth of superconducting phase is extremely slow. 
Further, we assume that the magnetic field strength at the core
region of a strange star are much less than the corresponding critical
value for the destruction of superconducting property and the temperature
is also low enough. Then during such a type-I superconducting phase
transition, the  magnetic  flux  lines  from the superconducting
quark matter of the strange star will be pushed out  towards  the
normal  crustal  region.  Now for a small type-I superconducting
laboratory sample placed in an external magnetic field less than the
corresponding critical value, the expulsion  of  magnetic  field  takes  place
instantaneously. Whereas in the bulk strange star scenario, the picture may 
be completely different. It may take several thousands of years for the 
magnetic flux lines to get expelled from the superconducting core. Which
further means, that the growth of superconducting phase in strange stars
may not be instantaneous. A simple
estimate shows that the expulsion time due to ohmic diffusion is $\sim
10^4$yrs \cite{R22}. It was shown by Chau using Ginzberg-Landau
formalism that the time for expulsion of magnetic lines of force
accompanied by the enhancement of magnetic field non-uniformities at the 
crustal region gets prolonged to $10^7$yrs \cite{R23}. Alford et. al. 
investigated the expulsion of magnetic lines of force from the colour 
superconducting region by considering the pairing of like and unlike quarks 
and obtained the expulsion time much larger than the age of the Universe 
\cite{R24}.

In our investigation of magnetic flux expulsion from growing superconducting 
core of a strange star, the idea of impurity diffusion in molten alloys or the 
transport of baryon numbers from hot quark matter soup to hadronic matter 
during quark-hadron phase transition in the early universe, expected to occur 
micro-second after big bang (the first mechanism is used by the material 
scientists and metallurgists \cite{R24a}, whereas the
later one is used by cosmologists working in the field of big bang
nucleo-synthesis \cite{R25,R25a}) are assumed. In the present section, we shall
further show the possibility of Mullins-Sekerka normal-superconducting 
interface instability \cite{R26a,R26b} in quark matter. This is generally 
observed in the case of solidification of pure molten 
metals at the solid-liquid interface, if there is a temperature gradient. 
The interface will always be stable if the  temperature  gradient  is positive,
otherwise it will be unstable. In alloys, the criteria for stable / unstable 
behaviour is more complicated. It is seen that, during solidification of an
alloy, there is a substantial change in the  concentration  ahead
of the interface. Here solute diffusion as well as the heat flow
effects must be considered simultaneously. The
particular  problem  we are going to investigate here is analogous to
solute diffusion during solidification of an alloy.

\section{A Formalism for Slow Nucleation}  
It  has  been  assumed  that  the growth of superconducting quark
bubble started from the centre of the star  and  the
nomenclature {\it{controlled growth}} for such phenomenon has been used. If the
magnetic  field strength and the temperature of the star are a few orders of 
magnitude less than their critical values, the normal quark matter phase is
thermodynamically unstable relative to the corresponding
superconducting one. Then due to fluctuation, a droplet of
superconducting quark matter bubble may be nucleated in metastable
normal quark matter medium. If the size of this superconducting bubble is 
greater than the corresponding critical value, it will act as the
nucleating  centre  for the growth of superconducting quark core.
The critical radius can be obtained by minimising the free energy.
Then following the work of Mullins and Sekerka, we have \cite{R26a,R26b}
\begin{equation}
r_c=\frac{16\pi\alpha}{B^{(c)2}\left [1-\left (\frac{B}{B^{(c)}}\right )^2
\right ]},
\end{equation}
where $\alpha$ is  the  surface tension or the surface energy per unit area of 
the critical superconducting bubbles, (from this expression it is possible to 
obtain the critical size of the quark matter bubble by considering $10^{-3}
\leq \alpha \leq 1$ (in MeV/fm$^{-3}$) as the range of surface tension) which 
is greater than zero for a type-I superconductor-normal  interface,  
$B^{(c)}$  is  the  critical magnetic field. In presence of a magnetic field 
$B< B^{(c)}$, the  normal  to  superconducting  transition  is  first  order in
nature. As the  superconducting  phase  grows  continuously,  the
magnetic  field  lines  will  be pushed out into the normal quark
matter crust. This is the usual {\it {Meissner effect}} observed
in type-I superconductor. We compare this phenomenon of  magnetic
flux  expulsion  from a growing superconducting quark matter core with the
diffusion of impurities from the frozen phase of molten metal or the transport 
of baryon numbers from hot quark matter soup during quark-hadron phase
transition in the early universe. The  formation  of
superconducting  zone  is  compared  with  the  solidification of
molten metal or with the transition to hadronic phase with almost zero baryon 
number. It is known from the simple thermodynamic  calculations
that  if the free energy of molten phase decreases in presence of
impurity atoms, then during solidification they prefer to  reside
in  the  molten phase otherwise they go to the solid phase. In this particular 
case the magnetic field lines play the role of impurity atoms and because of 
less free energy, they prefer to remain in normal quark matter phase. The 
normal quark  matter phase plays the role of molten metal or the hot quark
soup. Whereas the superconducting phase can be compared with the frozen solid 
phase or the hadronic phase. This idea was applied to baryon number transport 
during first order quark-hadron phase transition in the early Universe,
where baryon number replaces impurity, quark phase replaces molten  metal  and
hadronic  matter replaces that of solid metal \cite{R25,R25a}. Of course
the baryon number prefers to stay in the quark phase because of Boltzmann 
suppression factor in the hadronic phase. Since the magnetic flux lines prefer 
to reside in the normal phase, the well known Meissner effect can therefore be 
restated as {\it{the solubility of magnetic flux lines in the superconducting 
phase is zero  with a finite penetration depth}}.

The  dynamical  equation  for  the flux expulsion can be obtained
from  the  simplified  model  of   sharp   normal-superconducting
interface.  The  expulsion  equation  is  given by the well known
diffusion equation \cite{R27}
\begin{equation}
\frac{\partial B}{\partial t}=D\nabla^2 B
\end{equation}
where $B$ is the  magnetic  field  intensity  and  $D$  is  the
diffusion coefficient, given by
\begin{equation}
D=\frac{c^2}{4\pi \sigma_n},
\end{equation}
$\sigma_n$ is the electrical conductivity of the normal quark matter 
and $c$ is the velocity of light. The electrical conductivity of quark matter 
for $B=0$ is given by 
\cite{R28,R28a}
\[
\sigma_n \sim \alpha_s^{-3/2} T_{10}^{-2} 
\]
or \cite{D48}
\begin{equation}
\sigma_n\sim (\alpha_s T_{10})^{-5/3}
\end{equation}
expressed in sec$^{-1}$, where $\alpha_s$ is the strong coupling constant
and $T_{10}=T/10^{10}$K, the numerical value for this electrical
conductivity in  the  case  of  quark matter relevant for quark star density  
is  $\sim 10^{26}$ sec$^{-1}$. In the order of magnitude estimate, we have 
used this numerical value for electrical conductivity. In this context we must mention that in presence of 
strong quantizing magnetic field, $\sigma_n$ may not be a scalar quantity, 
magnetic field destroys the isotropy of the electromagnetic properties
of the medium. In particular, for extremely large $B$, the components 
of electric current vector orthogonal to $B$ become extremely small. Which 
indicates that the quarks can move only along the direction
of magnetic field or in other words, across the field the resistivity becomes 
extremely high. When the conditions for charge neutrality and $\beta$-equilibrium are
considered together, since the mass of $s$-quark is assumed to be $150$Mev, 
the electron density can not become exactly zero, but it is a few orders of 
magnitude less than the $s$-quark density. Therefore, one can neglect the 
electron contribution to  electrical conductivity. 

A solution of eqn.(2) with spherical symmetry can be obtained by Greens' 
function technique, and is given by (for a general topological structure, no
analytical solution is possible)
\begin{eqnarray}
B(r,t)&=&\frac{1}{2r (\pi D t)^{1/2}}\int_0^\infty B^{(0)}
(r')\nonumber \\ &&\big [ \exp(-u_-^2)-\exp(-u_+^2)\big ]r'dr'
\end{eqnarray}
where  $u_\pm=(r\pm  r')/2(Dt)^{1/2}$  and  $B^{(0)}(r)$ is the
magnetic field distribution within the star at $t=0$, which is of
course an entirely unknown function of radial coordinate $r$.  To
obtain an estimate of magnetic field diffusion time scale ($\tau_D$),
we assume $B^{(0)}(r)=B^{(0)}=$ constant. Then we have from eqn.(5)
\begin{equation}
B(r,t) =  B^{(0)}\left [ 1-\frac{2}{r} (\pi D t)^{1/2}\right ] 
\end{equation}
Hence, if we put $B(r,t)=0$ (field free condition), the estimated time scale 
for the expulsion of magnetic flux lines is $\sim 10^5-10^6$yrs. Which
is of the same order of magnitude as the Ohmic decay scale. From this 
simple estimate, this is the approximate time scale for complete expulsion 
of magnetic field lines. Which is of course quite large. Unfortunately, nothing else can 
be inferred about the growth of superconducting zone and the associated 
expulsion of magnetic flux lines from this region. The reason behind such 
uncertainty is our lack of knowledge or definite ideas on the numerical values 
of the parameters present in eqn.(5). 

To get some idea of the effect of magnetic field on the structure of growing 
superconducting zone, we shall now investigate the morphological instability of
normal-superconducting interface of  quark  matter. The  motion  of 
normal-superconducting  interface  is extremely important in this case and has 
to be taken into consideration. Then instead of eqn.(2) which is valid in the  
rest frame, an equation expressed in a coordinate system which is moving with 
an element of the boundary layer is the correct description of such 
superconducting growth, known as {\it{Directional Growth}}. The equation is 
called {\it{Directional Growth Equation}}, and is given by
\begin{equation}
\frac{\partial  B}{\partial  t}  -v  \frac{\partial B}
{\partial z} =D\nabla^2 B
\end{equation}
where  the motion of the plane interface is assumed to be along z-axis and 
$v$ is the velocity of  the  front.  This  diffusion  equation  must  be
supplemented  by  the  boundary  conditions at the interface. The
first boundary condition is obtained by  combining  Ampere's
and Faraday's laws at the interface, and is given by
\begin{equation}
B v\mid_s=-D(\nabla B) .\hat n \mid_s 
\end{equation}
where   $\hat   n$  is  the  unit  vector normal to the interface
directed from the normal phase to the superconducting phase. This
is nothing but the continuity equation for magnetic flux diffusion. For
the normal growth of superconducting zone, the  rate
at  which  excess  magnetic  field  lines  are  rejected from the
interior of the phase is balanced by the rate at  which  magnetic
flux lines diffuses ahead of the two-phase interface. Therefore the boundary  
layer  between superconducting-normal quark
matter phases will become unstable if excess magnetic field lines are present
on the surface  of  the  growing  superconducting  bubble, i.e., if the
rate of diffusion of magnetic flux lines is slow enough compared to the
rate at which they are expelled from the superconducting zone. Local
thermodynamic  equilibrium at the interface gives (Gibbs-Thompson
criterion)
\begin{equation}
B \mid_s  \approx  B^{(c)}  \left  (  1-\frac{4\pi   \alpha
}{RB^{(c)^2}}\right )
=  B^{(c)}  \left  (  1- \delta C \right )
\end{equation}
where  $\delta$  is  called  capillary  length  with $\alpha$ the
surface tension, $C$ is the curvature  $=1/R$  (for  a  spherical
surface), and $B^{(c)}$ is the thermodynamic critical field.

To investigate the stability of superconducting-normal interface,
we  shall  follow  the original work by Mullins and Sekerka \cite{R26a,R26b},
and consider a steady state growth of superconducting core. Then
the  time  derivative  in  eqn.(7)  will  not appear. Introducing
$r_\perp=(x^2+y^2)^{1/2}$ as the transverse  coordinate  at  the
interface, we have after rearranging eqn.(7)
\begin{equation}
\left      [\frac{\partial^2      }{\partial      r_\perp^2}+
\frac{1}{r_\perp}\frac{\partial}{\partial r_\perp}+
\frac{\partial^2 }{\partial z^2}+
\frac{v}   {D}\frac{\partial   }{\partial  z}\right  ]  B=0
\end{equation}
The approximation that the solidification is occurring under steady state condition
used in the freezing of molten material will be followed in the present case
of normal to superconducting phase transition. Now if it is
assumed that these two phenomena taking place in 
two completely separate physical world are almost
identical natural processes, then the concentration of magnetic flux 
lines and normal-superconducting interface morphology will be independent of
time. The main disadvantage of this assumption  is  that there will be no
topological evolution  of  the  interface shape. As a consequence of this
constraint the solution  to the basic  diffusion  problem
is   indeterminate   and   a   whole  range  of  morphologies  is
permissible from the mathematical point  of  view.  In  order  to
distinguish  the  solution which is the most likely to correspond
to reality, it is necessary to  find  some  additional  criteria.
The examination  of  the  stability  of a slightly perturbed growth
form is probably the most reasonable manner in which
this situation may be treated. In the following we shall investigate  the
morphological  instability  of  normal-superconducting  interface
from eqn.(10). Assuming a solution of this equation expressed as the product of
separate functions of $r_\perp$ and $z$ and setting the
separation  constant  equal  to  zero and  using the boundary
condition given by eqn.(9), we have for an  unperturbed  boundary
layer moving along $z$-axis
\begin{equation}
B=B^{(s)} \exp(-zv/D) =B^{(s)} \exp(-2z/l)
\end{equation}
where $l=2D/v$ is the layer thickness, which is  very  small  for
small $D$. Mathematically, the thickness of this layer is infinity.
For practical purpose an effective value $l$ can  be  taken.  The
order  of  magnitude  estimates  or limiting values for the three
quantities $D$, $v$ and $l$ can be obtained  from  the  stability
condition of planer interface.

Due to excess magnetic flux lines at the interface, the form of
the planer  normal-superconducting  interface  described  by  the
equation  $z=0$  is assumed to be changed by a small perturbation
represented by the simple sine function
\begin{equation}
z=\epsilon \sin(\vec k.\vec r_\perp)
\end{equation}
where  $\epsilon$  is  very  small  amplitude and $\vec k$ is the wave
vector of the perturbation. Then the perturbed  solution  of  the
magnetic field distribution near the interface can be written as
\begin{equation}
B=B^{(s)}\exp(-vz/D)      +A\epsilon     \sin(\vec     k.\vec
r_\perp)\exp(-bz)
\end{equation}
where  $A$  and $b$ are two unknown constants. Since the solution
should satisfy the diffusion equation (eqn.(10)), we have
\begin{equation}
b=\frac{v}{2D}+\left     [     \left    (\frac{v}{2D}\right
)^2+k^2\right ]^{1/2}
\end{equation}
To  evaluate  $A$,  we utilise the assumption that both $\epsilon$ and
$\epsilon \sin(\vec k.\vec r_\perp)$ are small enough so that  we
can  keep  only the linear terms in the expansion of exponentials
present  in  eqn.(13). Then at the interface, we have after some straight forward 
algebraic manipulation
\begin{equation}
A=\frac{v}{D} B^{(s)}
\end{equation}
The  expression  describing the magnetic field distribution ahead
of the slightly perturbed interface then reduces to
\begin{equation}
B=B^{(s)} \left [ \exp(-vz/D)  +\frac{v}{D}  \epsilon  \sin
(\vec k.\vec r_\perp)\exp(-bz) \right ]
\end{equation}

Now from the other boundary condition (eqn.(9)) we have
\begin{equation}
B^{(s)}=B^{(c)}-\frac{4\pi \alpha B^{(c)}}{B^{(s)2}}C
\end{equation}
where $C=z^{''}/(1+z^{'2})^{3/2}$ is the curvature at $z=\epsilon
\sin(\vec  k.\vec  r_\perp)$  and prime indicates derivative with
respect to $r_\perp$.

Neglecting $z^{'2}$, which is small for small perturbation, we have
\begin{equation}
B^{(s)}=B^{(c)} +\Gamma k^2 S
\end{equation}
where $\Gamma =4\pi  \alpha  B^{(c)}/B^{(s)2}$  and  we  have
replaced  $\epsilon  \sin(\vec k.\vec r_\perp)$ by $S$. Since the
amplitude of perturbation  $\epsilon$  is  extremely  small,  the
quantity $S$ is also negligibly small.

Now eqn.(18) can also be written as
\begin{equation}
B^{(s)}= B^{(c)}+ GS
\end{equation}
where
\begin{equation}
G=\frac{dB}{dz}\mid_{z=S}     =-\frac{v}{D}    \left    (
1-\frac{vS}{D} \right ) B^{(s)} -bAS (1-bS) 
\end{equation}
Combining these two equations, we have
\begin{equation}
k^2\Gamma +\frac{v}{D}  \left  (  1-\frac{vS}{D}  \right  )
B^{(s)} -\frac{bv}{D} B^{(s)} S(1-bS)=0
\end{equation}
This  expression determines the form  (values of $k$) which the
perturbed interface must assume in order to satisfy  all  of  the
conditions  of the problem. To analyse the behaviour of the roots,
we replace right hand side of eqn.(21) by  some  parameter  $-P$.
(We  have  taken  $-P$  in order to draw a close analogy with the
method given in refs.\cite{R25,R25a}). Then rearranging eqn.(21), we have
\begin{equation}
-k^2\Gamma  -\frac{v}{D}  \left  (  1-\frac{vS}{D} \right )
B^{(s)} +\frac{bvB^{(s)} S}{D} (1-bS) =P
\end{equation}
(in  refs.\cite{R25,R25a}  the  parameter  $P$  is  related  to  the time
derivative of $\epsilon$, the amplitude of small perturbation). If
the  parameter  $P$  is  positive  for  any  value  of  $k$,  the
distortion  of  the  interface  will  increase,  whereas, if it is
negative for all values of $k$, the perturbation  will  disappear
and  the interface will be stable. In order to derive a stability
criterion, it only needs to know whether eqn.(20) has roots  for
positive values of $k$. If it has no roots, then the interface is
stable  because  the  $P-k$  curve never rises above the positive
$k$-axis and $P$ is therefore negative for  all  wavelengths.  We
have  used Decarte's theorem to check how many positive roots are
there. It is more convenient to express $k$ in terms of  $b$  and
then replacing $b$ by $\omega +v/D$. Then we have from eqn.(22)
\begin{eqnarray}
-\omega^2 \left ( \Gamma +\frac{vB^{(s)} S^2}{D}  \right  )
&-& \omega  \left ( \Gamma +\frac{2vB^{(s)} S^2}{D} -B^{(s)} S
\right  )  \frac{v}{D}  \nonumber\\ &-&\frac{v}{D}  B^{(s)} 
\left  (  1- \frac{v}{D} S\right )^2 =P
\end{eqnarray}
This  is  a  quadratic equation for $\omega$. The first and the
third terms are always negative. The second  term  will  also  be
negative if
\begin{equation}
\Gamma + \frac{2vB^{(s)} S^2}{D} -B^{(s)} S>0
\end{equation}
Then  it follows from Decart's rule that if the condition (24) is
satisfied, there can not be any positive root. Which implies that
the small perturbation of the interface will disappear. Since the
amplitude of perturbation is assumed to be extremely  small,  the
quantity   $S=\epsilon   \sin(\vec  k  .\vec  r_\perp)$  is  also
negligibly small. Under such circumstances  the  middle  term  of
eqn.(23)  is  much  smaller  than rest of the terms. The Decart's
rule given by the condition (24) can be re-written as
\begin{equation}
\Gamma > B^{(s)} S
\end{equation}
Which after some simplification gives the stability criterion for
the plane unperturbed interface, given by
\begin{equation}
\alpha >\frac{B^{(s)3}S}{4\pi B^{(c)}}
\end{equation}

From   the   stability   criterion,   it   follows    that    the
normal-superconducting  interface energy/area of quark matter has
a lower bound, which depends  on  the  interface  magnetic  field
strength,  critical  field  strength and also on the perturbation
term $S$. An order  of  magnitude  of  this  lower  limit  can  be
obtained by assuming $B^{(s)}=10^{-3}B^{(c)}$.
(Since  the  critical  field  $B^{(c)}\sim  10^{16}$G,  and the
neutron star magnetic field strength $B \sim 10^{13}$G,  we  may
use this equality). Then the lower limit is given by
\begin{equation}
\alpha_L\approx  10^{-9}~~{\rm{MeV/fm}}^2~\left( \frac{S}{{\rm
{fm}}}\right )
\end{equation}
On the other hand for $B^{(s)}=0.1 B^{(c)}$, we have
\begin{equation}
\alpha_L\approx  10^{-3}~~{\rm{MeV/fm}}^2~\left( \frac{S}{{\rm
{fm}}}\right )
\end{equation}
The approximate general expression for the lower limit  is  given
by
\begin{equation}
\alpha_L\approx  h^3~{\rm{MeV/fm}}^2~\left( \frac{S}{{\rm
{fm}}}\right )
\end{equation}
where  $h=B^{(s)}/B^{(c)}$.  Therefore  the maximum value of
this lower limit is
\begin{equation}
\alpha_L^{\rm{max.}}\approx  1~{\rm{MeV/fm}}^2~\left( \frac{S}{{\rm
{fm}}}\right )
\end{equation}
when the two phase are in thermodynamic  equilibrium.  Of  course
for  such  a strong magnetic field, as we have seen \cite{R6,R7} that 
there can not be a first order quark-hadron phase  transition.

On  the  other  hand  if  we do not have control on the interface
energy, which can in principle be obtained  from  Landau-Ginzberg
model,   we  can  re-write  the  stability  criteria  in terms  of
interface concentration of magnetic field  strength  $B^{(s)}$,
and is given by
\begin{equation}
B^{(s)}  < \left [ \frac{4\pi \alpha B^{(c)} }{S\left ( 1-
\frac{2v}{D}S\right )} \right ]^{1/3}
\end{equation}
This  is more realistic than the condition imposed on the surface
tension $\alpha$. Now for a type-I  superconductor,  the  surface
tension  $\alpha  >0$,  which implies $1-2vS/D >0$. Therefore, we
have  $2vS/D  <  1$. For  the   typical   value of  $\sigma_n
\sim10^{26}$ sec$^{-1}$ for the electrical conductivity of normal
quark  matter, the profile velocity $v< D/2S \sim 10^{-6}/S$cm/sec 
$\sim 1$cm/sec for  $S\sim  10^{-6}$cm. Therefore the interface velocity 
$<1$cm/sec for such typical values  of $\sigma_n$ and $S$ to make the planer 
interface stable under small perturbation. Now the thickness of the layer  at  
the interface   is  $l=2D/v>10^{-6}$cm  for  such  values  of  $D$  (or
$\sigma_n$)  and  $v$.  Here  $S$  is  always  greater  than  $0$,
otherwise,     the     magnetic    field    strength    at    the
normal-superconductor interface becomes unphysical. As before, if
the second term of eqn.(23) is negligibly small compared to other
two terms, we have
\begin{equation}
B^{(s)}  <  \left  [  \frac{4\pi \alpha B^{(c)}}{S}\right
]^{1/3} 
\end{equation}
For the sake of illustration, we have shown in fig.(1), the distribution
of magnetic field in a very small portion of horizontal plane at the
perturbed interface. We have solved eqn.(17) numerically and use the typical
parameter set as given above. Along z-axis, we have plotted the ratio 
of the strength of surface magnetic field
and the critical strength for the destruction of type I quark matter
superconductivity and x and y axes represent the x-y coordinates of this tiny
horizontal plane. The distribution is extremely chaotic in nature and
some of the magnetic field peaks are stronger that the critical strength
to destroy type-I superconductivity. Then analogous to the case of
crystal growth in an impure molten alloy, where the process is stopped because of
high density of accumulated impurities at the interface, here also,
further growth of superconducting phase will be stopped. Therefore, it
is a kind of instability at the interface, which forces the growth
process to stop abruptly. Since the slow growth of superconducting phase
does not give a stable quark star with very low surface magnetic field, is
therefore not a physically acceptable phenomenon.

\section{The Prompt Process of Flux Expulsion}
We have noticed that the slow expulsion process, in which diffusion
mechanism of
magnetic lines of force may be applicable, leads to a kind of
instability at the interface. It ultimately stops abruptly the growth process 
of superconducting zone. As a consequence we will not get a stable quark
star with very low surface magnetic field.

In this section, we assume that the superconducting phase of quark
matter grows by a faster process; so that the diffusion model for the
magnetic lines of force is not applicable. Which actually
means, that the expulsion also occurs with a faster rate. Now during
this process, since magnetic flux changes rapidly with time, an emf will
be induced in the system. Although, the superconducting quark Cooper
pairs do not feel electromagnetic force, a Lorentz force will be exerted
on the electrons near the surface region, which are in normal phase.
These electrons will try to nullify the force following Lenz's law. As a
consequence they will start gyrating about the magnetic lines of force
at the surface / in the electro-sphere. We have already argued that these circulation of the electrons about the
magnetic field lines will produce a diamagnetic effect to oppose the
Lorentz force by reducing the strength of magnetic field at the surface.

To get an estimate for induced magnetic field strength at the strange
star surface / electro-sphere and the corresponding reduction in magnetic field strength, 
we follow the recent article by Hirsch \cite{H4}. The magnetic dipole moment 
per electron in the electro-sphere, due to the tiny orbital motion is given by
\begin{equation}
\mu= \frac{eu}{2} a ~~~~({\rm{we ~have ~assumed ~that}}~~c=1)
\end{equation}
where $u$ is the orbital velocity and $a$ is the radius of the orbit. The
induced emf in the electronic orbit is then given by
\begin{equation}
E=\frac{a}{2}\frac{\partial B}{\partial t}
\end{equation}
Then the corresponding change in orbital speed is given by
\begin{equation}
\Delta u=\frac{ea}{2m_e}B
\end{equation}
Hence the change in magnetic moment per electron is
\begin{equation}
\Delta \mu=\frac{e^2a^2}{4m_e}B,
\end{equation}
and the associated average value of the induced magnetization per unit volume
in the electro-sphere is given by
\begin{equation}
{\cal{M}}=\frac{n_ee^2a^2}{4m_e}B
\end{equation}
where $n_e$ is the average electron density in the electro-sphere
(in reality, $n_e$ must be maximum near the quark matter surface
and is minimum at the outer surface of the electro-sphere).
Hence one can obtain the strength of induced magnetic field at the surface
region, which is given by
\begin{equation}
B_{\rm{ind}}=4\pi \cal M
\end{equation}
Therefore the ratio of induced magnetic field by the electrons in the
electro-sphere to the existing strange star magnetic field is given by
\begin{equation}
\frac{B_{\rm{ind}}}{B}=\frac{n_ee^2}{m_e}\pi a^2 
\end{equation}
The strength of induced magnetic field is therefore a monotonically
increasing function of electron density. It is very easy to verify that
for complete suppression of magnetic field of a strange star, with a 
typical average electron density $n_e=10^{-5}n_B$, where
$n_B=0.17$fm$^{-3}$, the normal nuclear density, the orbital radius 
$a\sim 7$fm. Which is about three orders of magnitude less than the width of 
the electro-sphere. Therefore it is quite obvious that the electro-sphere of
width $\sim 1000$fm can accommodate a large number of tiny magnets. It
may 
be argued that the actual form of  
trajectories for such tiny magnets in each half of the electro-sphere are closed 
helical spring. 

Now, if there is no reduction, the magnetic field strength at the surface could be as high as 
$10^{30}$ times the magnetic field of a typical neutron star. One can obtain 
this number by considering the ratio $R^2/d^2$, where $R=10$Km, the radius of 
the star and $d=1000$fm, the width of the electro-sphere and assuming that the
magnetic flux remains conserved during Meissner expulsion from 
super-conducting quark matter. This value is far above the upper limit 
predicted by Shabad and Usov \cite{UV}.

Now it has been discussed in the literature that if the density of quark matter,
particularly, at the core region is sufficiently high, the 
color super-conducting quark matter undergoes a phase transition to 
what we call the CFL phase \cite{CF1,CF2,CF3}. It is therefore expected
that at the ultra-dense interior, there will be a color as well as charge neutral CFL 
phase. 
The number of quarks in the CFL cluster is even and divisible by three. It is 
found that in this new phase all the three flavors $u$, $d$ and $s$ can form 
pairs with the same flavor as well as with other components, i.e., their
Fermi 
levels are identical. In this case the excess $u$-quarks will 
therefore go to the 
outer region. The gross structure as mentioned at the 
beginning of this article may not therefore be correct for a strange
star with extremely high core density; at the interior, it 
could be the CFL phase, at the outer core or inner crust region, it is the 
positively charged color neutral quark matter in non-CFL color super-conducting 
phase and finally the thin outer surface, the electro-sphere, is the normal 
electron gas.
Whether the formation of CFL phase at the inner part of a super-conducting
strange star is an equivalent QCD picture of the so called Tau-effect 
\cite{TAU1,TAU2} needs further investigation.

Therefore we expect that depending on the density of quark matter, the inner 
part of a strange star is either a type-I super-conductor or in CFL phase, the
outer region is color neutral type-I super-conducting non-CFL phase,
mainly dominated by $u$-quarks and some electrons, and
finally the
outer surface is a thin layer of electron gas in normal state. The magnetic
field lines are expelled from quark matter phase to the electron gas layer
with a small penetration at the outer surface of the quark matter.
We have conjectured that
the classical diamagnetism of the normal electron gas suppresses the magnetic 
field of a strange star by diluting the existing magnetic flux lines in the
electro-sphere by the repulsive action of the induced magnetic field.
If $B\approx B_{\rm{ind}}$, we have almost total suppression. In this
ideal situation, so far the emission due to
magnetic 
activities are concerned, the object becomes dark to the observers. However, 
because of vacuum instability, non-thermal surface $\gamma$ emission is 
possible. Since $Z$ is very high, the Coulomb field at the quark matter surface 
will be extremely strong, then $e^+e^-$ pairs may be produced from
vacuum by Schwinger mechanism \cite{SCHW}. The created $e^+$ gets
annihilated with one of the electrons in the electro-sphere and $e^-$ will 
occupy one of the empty energy levels. Therefore, $\gamma$ emission will also be 
forbidden if the star is extremely cold, when there will be no more vacant 
states for the produced electrons \cite{UVN}.

An alternative explanation for low magnetic field of strange stars can
also be obtained from a very simple model of magnetic circuits. The
combination of electro-sphere and the magneto-sphere may be treated as
an equivalent magnetic circuit with varying reluctance. Of course, there
are some open flux lines goes out after cutting the light cylinder. In
that case, we may assume that the whole universe is a complicated
network of magnetic circuits with varying reluctance and the strength of
sources of magneto motive force. From the
definition, we have total flux
\begin{equation}
N=\frac{\rm{MMF}}{R}
\end{equation}
where MMF is the effective magneto motive force and $R$ is the
reluctance, which is given by
\begin{equation}
R=\oint \frac{dl}{\alpha \mu}
\end{equation}
where $dl$ is the length of some flux carrying element of cross section
$\alpha$ and magnetic permeability $\mu$. Let us consider the passage of
expelled magnetic flux lines through surface region, when there is no gyrating
motion of electrons. Here we may take $\mu \approx 1$, the reluctance is then
given by.
\begin{equation}
R_1=\int \frac{dl}{\alpha}{\Big\vert}_{es}+R_{ms}
\end{equation}
where {\bf{es}} indicates the electro-sphere and {\bf{ms}} is the
magneto-sphere.
On the other hand, with the
gyrating electrons, since there is a diamagnetic effect at the surface /
electro-sphere,
we must have the permeability $\mu < 1$ in this region. In Gaussian
unit, we have 
$\mu=1-4\pi \chi$. Where from eqn.(37) the susceptibility is given by
\begin{equation}
\chi=\frac{\partial {\cal{M}}}{\partial B}=\frac{n_ee^2a^2}{4m_e}
\end{equation}
The reluctance is then given by
\begin{equation}
R_2=\int \frac{dl}{\mu \alpha^\prime}{\Big\vert}_{es}+R_{ms}
\end{equation}
where $\alpha^\prime$ is now the new cross sectional area of the flux
carrying element.
Since the MMF source and the total flux does not change by the gyration
of electrons at the surface, we have effectively
\begin{equation}
R_1=R_2
\end{equation}
which gives
\begin{equation}
\alpha^\prime=\alpha \frac{1}{\mu}
\end{equation}
Since $\mu \ll 1$, it is therefore quite obvious that the area of cross section of the
electro-sphere will increase by several orders of magnitude. As a result
the magnetic flux density will also be reduced by the same factor.
Now it is well known that Larmor radius $a\sim B^{-1}$, here $B$ is the
surface magnetic field. Therefore it is quite possible in the parameter
space (electron density and the corresponding Larmor radius) for the physically
acceptable values for electron density and Larmor radius, the surface
magnetic field can be as low as $10^8$G. For the typical values of
electron density $n_e\sim 10^{-12}n_0$ and Larmor radius $a\sim 10$ \AA,
the magnetic permeability becomes $\sim 0.012$. Which gives surface
magnetic field a few times $10^7$G. Since the Larmor radius increases
with the decrease in magnetic field strength, it will further increase
the induced magnetic field (see eqn.(39)), if it dominates over the
electron density. The induced magnetic field, which is diamagnetic in
nature will further reduce the existing field. So the solution must be 
self-consistent in nature. We believe that ultimately a steady state will be 
established in the electro-sphere. This is true for both the models.

\section{Conclusion}
From this investigation, we may conclude in a straight forward way that
if strange stars really exists with very low surface magnetic field, the
type-I superconducting transition in the quark matter phase must have occurred
within the star. Further, the transition process must be very prompt, so
that diffusion model for the expulsion of magnetic flux lines is not 
applicable, where the later gives rise to some kind of instability at the
interface and finally stops the growth of superconducting phase
abruptly. Further, the slow transition model does have any mechanism to reduce
the surface field strength, which is $\leq 10^8$G. 

Our next conclusion is that the transition is quite fast. The classical
diamagnetism of gyrating electrons at the surface / electro sphere may
be the possible source of opposing field to reduce the expelled surface
magnetic field by several orders of magnitude. It is quite obvious that
with this prompt transition model, using the idea of Hirsch, one can
obtain a surface magnetic field for the expected quark stars as low as
$10^8$G.

\newpage
\begin{figure}[ht]
\psfig{figure=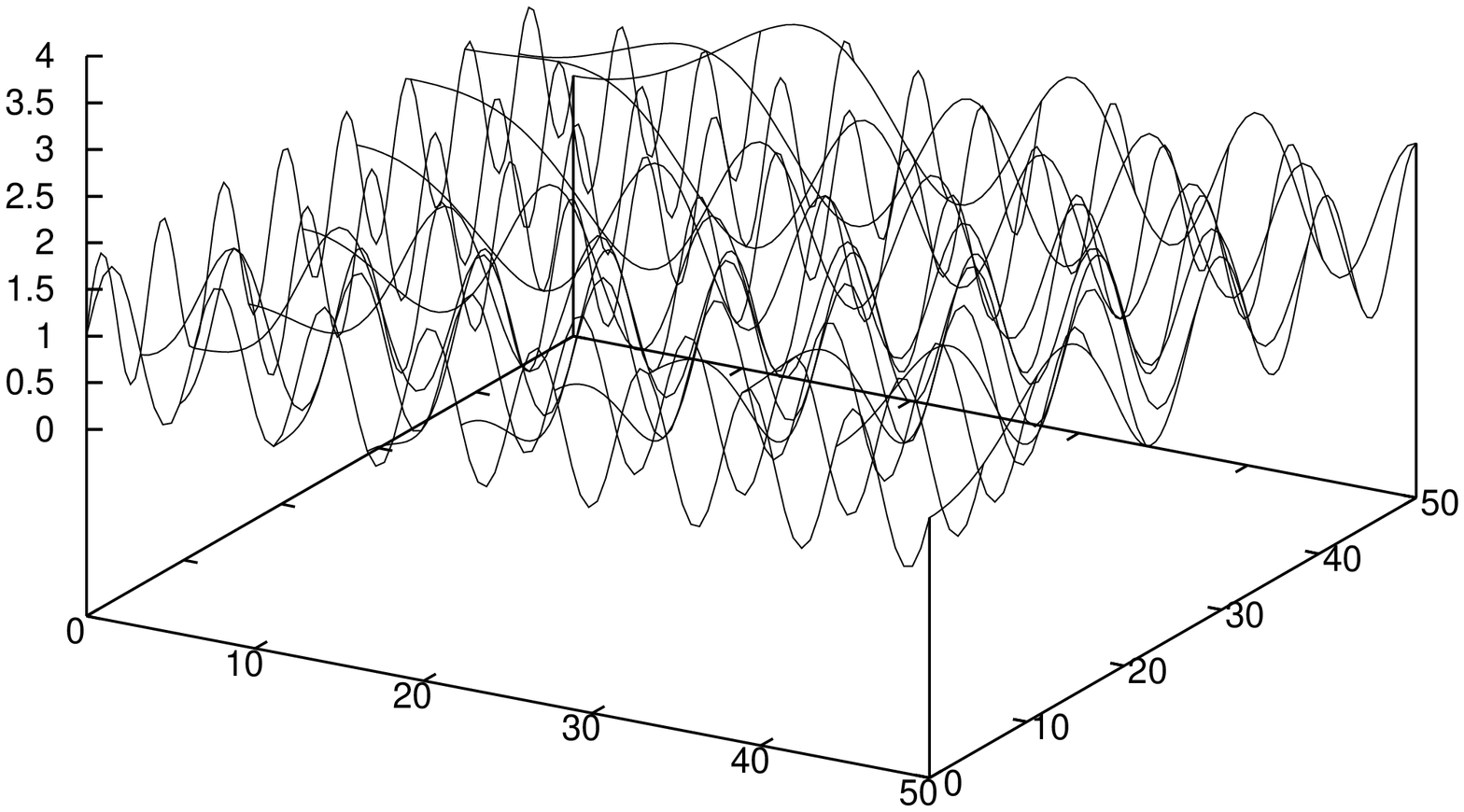,angle=270,height=0.4\linewidth}
\caption{The variation of magnetic field along $z$-axis (plotted as a
ratio of $B/B^{cr}$, where $B^{cr} \sim 10^{16}$G, the critical
strength to destroy the type-I quark matter superconductivity) with the
orthogonal coordinates $x$ and $y$.}
\end{figure}
\end{document}